\documentclass[prl,twocolumn]{revtex4}
\usepackage{epsfig}
\usepackage{bm}
\usepackage{amsmath}

\begin{document}

\title{Thermal convection, ensemble weather forecasting and distributed chaos}

\author{A. Bershadskii}

\affiliation{
ICAR, P.O. Box 31155, Jerusalem 91000, Israel}

\begin{abstract}
  Results of direct numerical simulations have been used to show that intensive thermal convection in a horizontal layer and on a hemisphere can be described by the distributed chaos approach. The vorticity and helicity dominated distributed chaos were considered for this purpose. Results of numerical simulations of the Weather Research and Forecast Model (with the moist convection and with the Coriolis effect) and of the Coupled Ocean-Atmosphere Mesoscale Prediction System (COAMPS) were also analysed to demonstrate applicability of this approach to the atmospheric processes. The ensemble forecasts of the real winter storms in the East Coast and Pacific Northwest as well as results of a simulation experiment with the multiscale storm-scale ensemble forecasts for eleven cases of mid-latitude convection in the central U.S. have been also discussed in this context.

\end{abstract}

\maketitle

\section{Distributed chaos}

  Systems with chaotic dynamics often have frequency power spectra with exponential decay \cite{fm}-\cite{mm}. For the systems described by dynamical equations with partial derivatives (in particular for the systems based on the Navier-Stokes equations) observations are less conclusive, especially for the wavenumber (spatial) power spectra. Figure 1 shows kinetic energy spectrum for a perturbation in statistically stationary isotropic homogeneous turbulence at Reynolds number $Re \simeq 2500$ \cite{bh} (the spectral data can be found at the site Ref. \cite{data}). In this paper a direct numerical simulation (DNS) with the Navier-Stokes equations   
$$   
\frac{\partial \bm u(\bm x,t)}{\partial t} + (\bm u\cdot \nabla) \bm u= -\nabla p+\nu\Delta \bm u   + {\bf f}  \eqno{(1)}
$$
$$
\nabla\cdot \bm u(\bm x,t)=0  \eqno{(2)}
$$    
was performed and a velocity field realization $\bm{u}_1$ was transformed into a new realization ${\bf u}_2$ by a slight instant perturbation of the forcing ${\bf f}({\bf x},t)$. Power spectrum of the field $\delta \bm{u} = \bm{u}_1 - \bm{u}_2$ was then computed as
$$
E_d(k,t) = \frac{1}{2} \int_{|\bm{k}|=k} d \bm{k} 
|\bm{\hat{u}}_1 ( \bm{k},t) -  \bm{\hat{u}}_2 ( \bm{k},t)|^2 \  \eqno{(3)}
$$
for a steady state.

   The dashed straight line in the Fig. 1 indicates the exponential decay
$$
E(k) = a \exp-(k/k_0)   \eqno{(4)}
$$  
 The insert to the Fig. 1 has been added in order to show that the $k_0$ from the Eq. (4) corresponds to the peak of the $E_d(k)$ spectrum. This is an indication of a tuning of the high-wavenumber chaotic dynamics to the coherent structures with the scale $k_0$. \\
\begin{figure} \vspace{-1.5cm}\centering
\epsfig{width=.45\textwidth,file=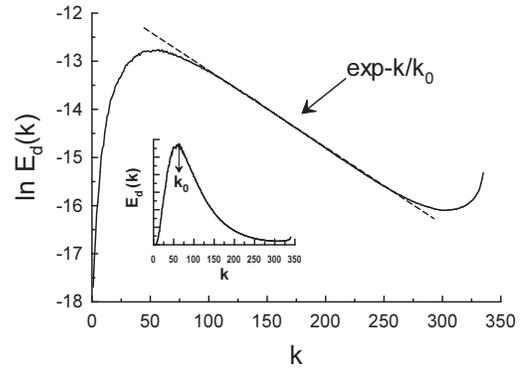} \vspace{-4cm}
\caption{Perturbation kinetic energy spectrum for the steady isotropic turbulence. The dashed straight line indicates the exponential decay Eq. (4).} 
\end{figure}

  Ensemble weather forecasting allows to take into account the intrinsic uncertainty in numerical forecasts of chaotic systems. In recent paper Ref. \cite{wd} results of an idealized ensemble simulation of mesoscale deep-convective systems were reported. A nonhydrostatic cloud-resolving model was used in order to generate ensembles of 20 perturbed and 1 control members. The ensembles were initialized by large-scale (91-km-wavelength) moisture perturbations with random phases. A strong line of thunderstorms was developed in all cases (see Ref. \cite{wd} for more details of the model configuration and simulation strategy).

   Figure 2 shows vertically averaged over the layer
$0 \leq z \leq 16$ km background (total) kinetic energy spectrum at 6 hours of the system development with 1km resolution (simulations were performed in a doubly periodic horizontal square domain of 512km $\times$ 512km, $k$ is horizontal wavenumber). The dashed curve indicates the exponential spectral decay Eq. (4) in the log-log scales (here and in all other figures $\log k \equiv \log_{10} k$). The faint straight line, indicating the '-5/3' slope in the log -log scales, is drawn in the figure for reference. Figure 3 shows corresponding vertically and ensemble averaged spectrum of perturbations in kinetic energy about the ensemble mean at the 6 hours of the system development. The dashed curve indicates the exponential spectral decay Eq. (4) in the log-log scales. The spectral data for the Figs. 2 and 3 were taken from the Fig. 7 of the Ref. \cite{wd}.\\ 
 
     In the general case of a statistical ensemble defined by parameters $a$ and $k_0$ the ensemble averaged spectrum can be represented by
$$
E(k) = \int P(a,k_0) ~\exp-(k/k_0)~ dadk_0  \eqno{(5)}
$$      
with a joint probability distribution $P(a,k_0)$. If the variables $a$ and $k_0$  are statistically independent, then
$$
E(k) \propto \int P(k_0) ~\exp-(k/k_0)~ dk_0  \eqno{(6)}
$$
with distribution $P(k_0)$ of the parameter $k_0$.\\

  Let the characteristic velocity $u_0$ vary with the scale $k_0$ in a scale invariant form (scaling)
$$  
u_0 \propto k_0^{\alpha}  \eqno{(7)}
$$
If the vorticity ${\boldsymbol \omega} ({\bf x},t)$ correlation integral
$$
I_{\omega} = \int \langle {\boldsymbol \omega} ({\bf x},t) \cdot  {\boldsymbol \omega} ({\bf x} + {\bf r},t) \rangle_V~  d{\bf r} \eqno{(8)}
$$
($<...>_V$ denotes the ensemble-volume average, cf. Ref. \cite{b2}) dominates the scaling Eq. (7), then from the dimensional considerations one obtains
$$
u_0 \propto I_{\omega}^{1/2} k_0^{1/2}  \eqno{(9)}
$$
   
\begin{figure} \vspace{-2cm}\centering
\epsfig{width=.42\textwidth,file=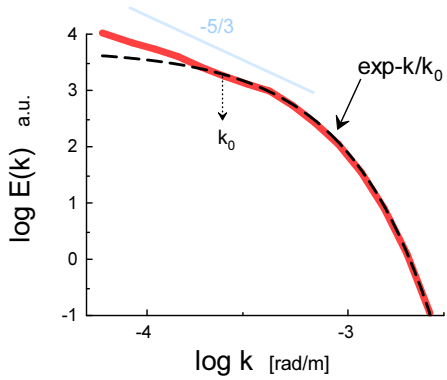} \vspace{-3.8cm}
\caption{Vertically and ensemble averaged background (total) kinetic energy spectrum at 6 hours of the system development (here and in all other figures $\log k \equiv \log_{10} k$).} 
\end{figure}
\begin{figure} \vspace{-1.5cm}\centering
\epsfig{width=.42\textwidth,file=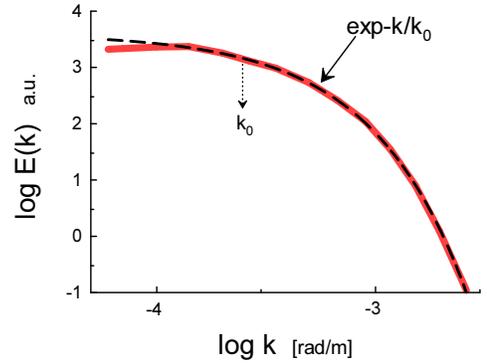} \vspace{-4.3cm}
\caption{As in Fig. 2 but for perturbation.} 
\end{figure}
  For Gaussian distribution of the characteristic velocity $u_0$ the variable $k_0$ has the chi-squared ($\chi^{2}$) distribution:
$$
P(k_0) \propto k_0^{-1/2} \exp-(k_0/4k_{\beta})  \eqno{(10)}
$$
here $k_{\beta}$ is a constant. 

   Substituting the Eq. (10) into the Eq. (6) one obtains
$$
E(k) \propto \exp-(k/k_{\beta})^{1/2}  \eqno{(11)}
$$

\section{Thermal convection}

At thermal (Rayleigh-B\'{e}nard) convection a horizontal layer of the fluid is cooled from top and heated from below. The Boussinesq approximation of the nondimensional equations describing the thermal convection is
$$
\frac{1}{\mathrm{Pr}}\left[ \frac{\partial \bf u}{\partial t} + (\bf u \cdot \nabla) \bf u \right] = -\nabla \sigma + \theta \hat{z} + \frac{1}{\sqrt{\mathrm{Ra}}} \nabla^2 \bf u, \eqno{(12)} 
$$
$$
\frac{\partial \theta}{\partial t} + (\bf u \cdot \nabla) \theta  =  u_z + \frac{1}{\sqrt{\mathrm{Ra}}} \nabla^2 \theta ,   \eqno{(13)}
$$
$$
\nabla \cdot \bf u  =  0,   \eqno{(14)}
$$
where $Pr$ is the Prandtl number, $Ra$ is the Rayleigh number, $\hat{z}$ is the buoyancy direction, and $\theta$ is deviation of temperature from the heat conduction state \cite{ssv}. \\

  Figure 4 shows kinetic energy spectrum computed for a direct numerical simulation of the thermal (Rayleigh-B\'{e}nard) convection at $Pr = 10^2$ and $Ra = 10^7$ (the spectral data for this figure were taken from Fig. 10 of the Ref. \cite{pvm}). The direct numerical simulation (DNS) was performed in a three-dimensional box with standard periodic boundary conditions on the lateral boundaries. On the bottom and top boundaries isothermal conditions for the temperature and free-slip conditions for velocity were used. The dashed curve in the Fig. 4 indicates the stretched exponential spectrum Eq. (11). \\

 Figure 5 shows kinetic energy spectrum computed for the Weather Research and Forecast Model \cite{ska} numerical simulation of the atmospheric moist convection without the Coriolis effect (the spectral data were taken from Fig. 10 of the Ref. \cite{srz}). Seven warm bubbles were used in the initial condition in order to initiate convection. The bubbles interact with each other under a wind shear (for more details see the Ref. \cite{srz}). The spectrum was averaged between 0 and 15 km of the height and over 4-6 hours of the evolution. The dashed curve indicates the stretched exponential spectrum Eq. (11).

\begin{figure} \vspace{-2.2cm}\centering
\epsfig{width=.45\textwidth,file=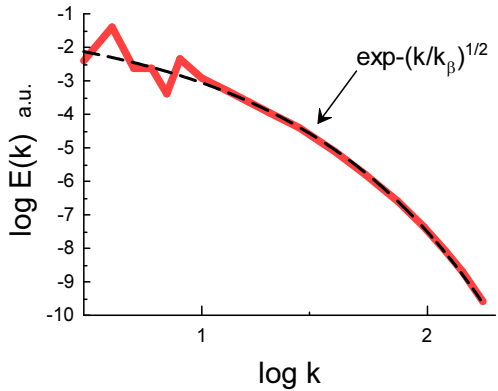} \vspace{-3.8cm}
\caption{Kinetic energy spectrum in the thermal (Rayleigh-B\'{e}nard) convection for $Ra = 10^7$ and $Pr = 10^2$ ( here and in all other figures $\log k \equiv \log_{10} k$).} 
\end{figure}
\begin{figure} \vspace{-0.5cm}\centering
\epsfig{width=.42\textwidth,file=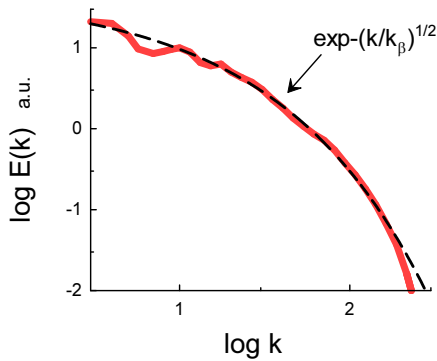} \vspace{-4cm}
\caption{Kinetic energy spectrum for the Weather Research and Forecast Model numerical simulation of the atmospheric moist convection.} 
\end{figure}
\begin{figure} \vspace{-1.5cm}\centering
\epsfig{width=.42\textwidth,file=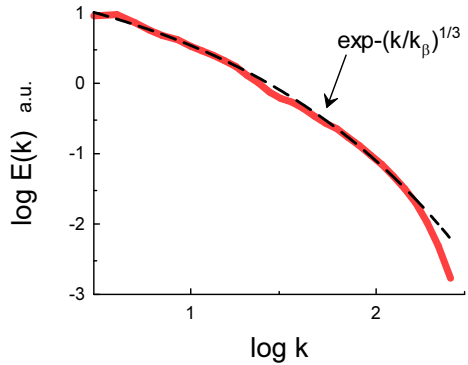} \vspace{-3.7cm}
\caption{As in Fig. 5 but with addition of the Coriolis effect.} 
\end{figure}
\section{Helicity dominated distributed chaos}

  The vorticity dominated thermal convection (distributed chaos) has the stretched exponential kinetic energy spectrum spectrum Eq. (11) (see also Ref. \cite{b3}). Therefore, let us look at a generalization:
$$    
E(k) \propto  \int P(k_0) ~\exp-(k/k_0)~ dk_0  \propto \exp-(k/k_{\beta})^{\beta}  \eqno{(15)}
$$
 If distribution of the characteristic velocity $u_0$ is $\mathcal{P} (u_0)$, then 
$$
\mathcal{P} (u_0) du_0 \propto P(k_0) dk_0 \eqno{(16)}
$$
Form the Eqs. (7) and (16) one obtains
$$
P(k_0)  \propto k_0^{\alpha -1} ~\mathcal{P} (u_0(k_0)) \eqno{(17)}
$$ 
  From the Eq. (15) asymptote of $P(k_0)$ at $k_0 \rightarrow \infty$ can be estimated as \cite{jon}
$$
P(k_0) \propto k_0^{-1 + \beta/[2(1-\beta)]}~\exp(-bk_0^{\beta/(1-\beta)}) \eqno{(18)}
$$
with a constant $b$.

  Then it follows from the Eqs. (7),(17) and (18) that for the Gaussian distribution $\mathcal{P} (u_0)$ the parameters $\alpha$ and $\beta$ are related by the equation
$$
\beta = \frac{2\alpha}{1+2\alpha}   \eqno{(19)}
$$
  
  For the helicity $h = ({\bf u} \cdot {\boldsymbol \omega})$ dominated distributed chaos the helicity correlation integral
$$   
I_h = \int  \langle  h ({\bf x},t) \cdot   h ({\bf x} + {\bf r},t) \rangle_V d{\bf r}  \eqno{(20)}
$$
should be used instead of the vorticity correlation integral. The helicity correlation integral $I_h$ was for the first time considered in the paper Ref. \cite{lt} and is known as the Levich-Tsinober invariant. It is usually associated with the helical waves \cite{l}. 

  Then it follows from the dimensional considerations:
$$
u_0 \propto I_{h}^{1/4} k_0^{1/4}  \eqno{(21)}
$$  
and using the Eq. (19) one obtains $\beta =1/3$, i.e.
$$
E(k) \propto \exp-(k/k_{\beta})^{1/3}  \eqno{(22)}
$$

   Figure 6 shows kinetic energy spectrum computed for the Weather Research and Forecast Model \cite{ska} numerical simulation of the atmospheric moist convection with the Coriolis effect (the spectral data were taken from Fig. 11a of the Ref. \cite{srz}). The dashed curve in the Fig. 6 indicates the stretched exponential spectrum Eq. (22) in the log-log scales (cf. previous Section, Fig. 5). 
   
  Figure 7 shows kinetic energy spectrum computed for a DNS of a Rayleigh-B\'{e}nard-like (thermal) convection on a hemisphere (the spectral data were taken from Fig. 18 of the Ref. \cite{bru} for the stationary state spectrum). The fluid was heated at the equator and the temperature gradient between the equator and the pole produces thermal plumes near the equator which move up toward the pole and initiate a thermal convection. 

\begin{figure} \vspace{-1.5cm}\centering
\epsfig{width=.42\textwidth,file=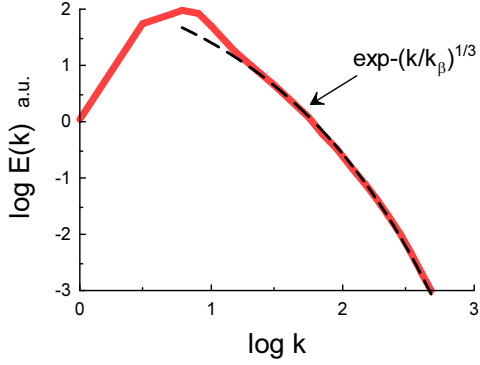} \vspace{-3.9cm}
\caption{Kinetic energy spectrum for the stationary state of the thermal convection on a hemisphere.} 
\end{figure}
\begin{figure} \vspace{-0.5cm}\centering
\epsfig{width=.42\textwidth,file=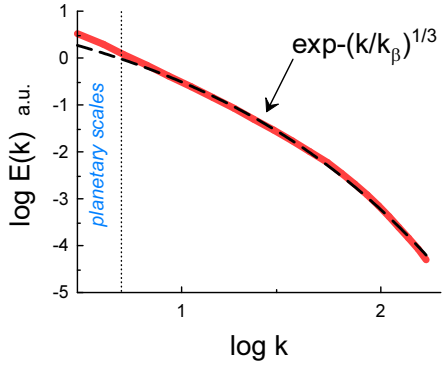} \vspace{-3.6cm}
\caption{Mean spectrum of kinetic energy in short-range weather forecasts (48-hours) experiment at 500 hPa.} 
\end{figure}
\begin{figure} \vspace{-1.6cm}\centering
\epsfig{width=.42\textwidth,file=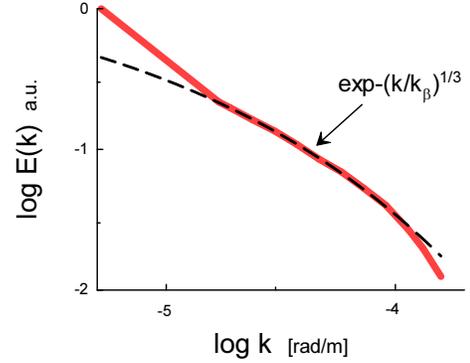} \vspace{-3.8cm}
\caption{Background kinetic energy spectrum for the 25 Dec. 2010 snowstorm.} 
\end{figure}
  The dashed curve in the Fig.7 indicates the stretched exponential spectrum Eq. (22) in the log-log scales. \\
  
  Figure 8 shows mean spectrum of kinetic energy in 48h weather forecasts experiment at 500 hPa. The spectral data were taken from the Fig. 7b of the Ref. \cite{buh} (the forecasts were made with the Environment Canada Deterministic Weather Forecasting Systems based on ensemble-variational data assimilation). The dashed curve indicates the stretched exponential spectrum Eq. (22) and covers Meso- and Synoptic scales (the dotted vertical line indicates the Planetary scales).

\section{Ensemble weather forecasting}

An ensemble forecast for an East Coast snowstorm was reported in Ref. \cite{dg}. 
The 100-member ensembles were generated by ensemble Kalman filter \cite{wh2}. The Coupled Ocean-Atmosphere Mesoscale Prediction System - COAMPS \cite{hod} was then used in order to integrate the ensembles for 36 hours forecast. The initial conditions were slightly altered for this purpose. The forecasting simulation started at 1200UTC 25 Dec. 2010 with real atmospheric data.
\begin{figure} \vspace{+0.3cm}\centering
\epsfig{width=.42\textwidth,file=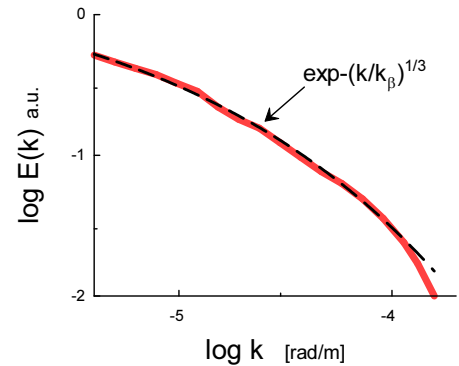} \vspace{-4.35cm}
\caption{Kinetic energy spectrum of the perturbation for the 25 Dec. 2010 snowstorm at 36 hours of the lead time.} 
\end{figure}
   
\begin{figure} \vspace{-1.5cm}\centering
\epsfig{width=.42\textwidth,file=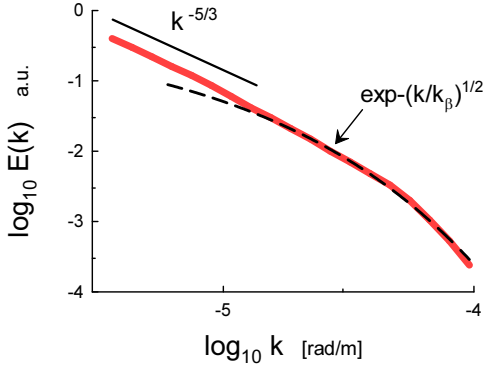} \vspace{-4cm}
\caption{Mean horizontal kinetic energy spectrum at the height
700-hPa at 1200UTC 17 Dec. 2008.} 
\end{figure}
\begin{figure} \vspace{-0.5cm}\centering
\epsfig{width=.42\textwidth,file=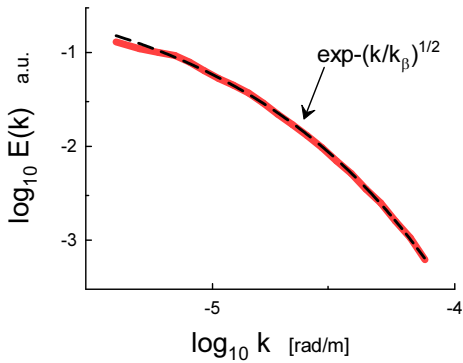} \vspace{-3.86cm}
\caption{Kinetic energy spectrum of the perturbations for the 17 Dec. 2008 snowstorm at the 36 hours of the lead time.} 
\end{figure}
\begin{figure} \vspace{-0.5cm}\centering
\epsfig{width=.42\textwidth,file=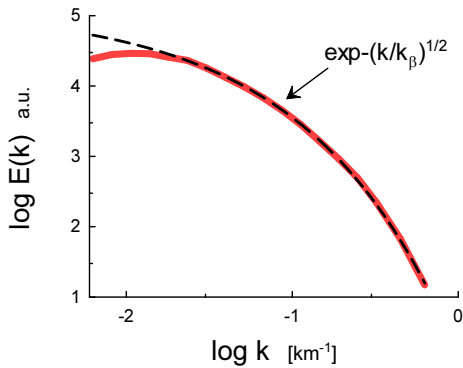} \vspace{-3.7cm}
\caption{Power spectrum of ensemble perturbations: ensemble member minus ensemble mean (averaged over all ensemble and case members),  for the $u$ component of wind at 900 hPa for 3h forecast time.} 
\end{figure}
   Figure 9 shows the ensemble and meridional averaged kinetic energy spectrum  at the height 500 hPa. Figure 10 shows ensemble and meridional averaged kinetic energy spectrum of the initially generated perturbation at the 36 hours of the lead time (the data for the both figures were taken from Fig. 6b Ref. \cite{dg}).   The perturbation is the difference between one ensemble member and the ensemble mean. The dashed curves in the figures indicate the stretched exponential decay Eq. (22). The authors of the Ref. \cite{dg} believe that the perturbation growth in their simulation is a result of quasi-uniform amplification of the perturbation at all wavenumbers (see also Refs. \cite{dg},\cite{drd}-\cite{map}).

   Another snowstorm was studied by the same method for the Pacific Northwest in the Ref. \cite{drd}. Figure 11 shows the mean horizontal kinetic energy spectrum at the hight 700-hPa at 1200UTC 17 Dec. 2008 (the data were taken from Fig. 13 Ref. \cite{drd}). Figure 12 shows the kinetic energy spectrum of the initially generated perturbation at the same height at the 36 hours of the lead time (the data were taken from the Fig. 14d Ref. \cite{drd}). The forecasting simulation started at  0000UTC 17 Dec. 2008 with real atmospheric data. The dashed curves in the figures 11 and 12 indicate the stretched exponential decay Eq. (11). \\
   
   Finally, let us consider results of a simulation experiment with eleven cases of mid-latitude convection in the
central US \cite{jw}. In this experiment influence of the multiscale perturbations generated by initial conditions on the storm-scale ensemble forecasts was studied using the Weather Research and Forecasting Advanced Research Model and the Global Forecast System Model at NCEP (see for more details about the cases, configuration and simulation strategy in the Refs. \cite{jw},\cite{jon15}).  

    Figure 13 shows power spectrum of ensemble perturbations: ensemble member minus ensemble mean (averaged over all ensemble and case members),  for the $u$ component of wind at 900 hPa for 3h forecast time. The spectral data were taken from Fig. 2 of the Ref. \cite{jw}. The dashed curve indicates the stretched exponential decay Eq. (11).

\section{Discussion}

In the paper Ref. \cite{L69} a two-dimensional barotropic vorticity model with the scaling kinetic energy spectra $E(k) \propto k^{-5/3}$ and  $E(k) \propto k^{-7/3}$ was used in order to estimate predictability properties of the atmospheric phenomena. A vast amount of studies was then devoted to the multiscale systems' predictability for the cases with power-law (scaling) kinetic energy spectra (see, for instance, recent Refs. \cite{wd},\cite{srz} and references therein). The power-law spectra are related to the scale-local interactions (such as cascades, for instance) \cite{my}, whereas the exponential spectra are a result of the non-local interactions directly relating very different scales \cite{b4}. This difference has serious consequences for predictability \cite{b3}. The non-local interactions, directly relating large scales with small ones, provide a basis for more efficient predictability.

The above considered examples show that the distributed chaos approach with the stretched exponential spectra Eq. (15) seems to be more relevant for description of the the buoyancy driven fluid dynamics and, especially, for the ensemble weather forecasting \cite{v}. \\

\section{Acknowledgement}

I thank A. Berera and R.D. J. G. Ho for sharing their data and discussions, and S. Vannitsem for comments.

\end{document}